# IPHC Emittance-meters: Design and Development

**E. Bouquerel**[a,*]**, C. Maazouzi**[a]

[a] *IPHC, Université de Strasbourg, CNRS-IN2P3, F-67037 Strasbourg, France*

*E-mail*: Elian.Bouquerel@iphc.cnrs.fr

ABSTRACT: The *Institut Pluridisciplinaire Hubert Curien* (IPHC) of Strasbourg, which celebrates its 15$^{th}$ year in 2021, is composed of four departments. Each of these departments comes from a different scientific horizon such as eco-physiology, chemistry, subatomic research and medical imaging. IPHC was created with the ambition of having different competences to develop high-level multidisciplinary programs with the basis of scientific instrumentation. Beam diagnostics is one of the main fields that has been intensively investigated during all these years within the team of the Instrumentation of Accelerators. This paper focuses on one of its major achievements, the Allison emittance-meter, developed in the framework of SPIRAL2, MYRRHA and FAIR projects.

KEYWORDS: Emittance-meter; Diagnostics; Allison scanner.

---

[*] Corresponding author.

## 1. Introduction

The *Institut Pluridisciplinaire Hubert Curien* (IPHC) is an IN2P3 (*Institut National de Physique des Particules*) unit which is part of CNRS (*Centre National de Recherche Scientifique*). Also supervised by the University of Strasbourg, IPHC is located in an ideal environment for the development of cutting-edge technologies, mainly in the fields of chemistry and physics.

IPHC was born from the unification of an institute of subatomic physics, a laboratory of analytical chemistry and a center of ecophysiology. These three units have chosen to combine their skills and their means in 2006 and were recently joined by another a fourth one dedicated to the field of radiobiology and hadron therapy. By using each other's knowledge and tools, the nearly 400 people from the IPHC develop collaborations between these different departments with the aim to construct projects at the frontier of their disciplines and thus better answer to the new stakes of our society.

Long before all this time, research activities in Strasbourg were mainly based on accelerators. Indeed, from 1948, the physicists of the *Institut de Recherches Nucléaires* (name of the institute at that time) could rely on a 1.5 MeV Crockcroft-Walten accelerator. Then in 1959, a 5.5 MeV HVEC Van de Graaff followed by a 2-3 MeV HVEC were installed at the newly constructed *Centre de Recherches Nucléaires*. Later, a MP-10 Tandem (named VIVITRON) was built and used for more than 10 years from 1993 [1].

This half-century of research using accelerators has contributed to the emergence of new areas of expertise that have grown and refined over the years within the research and technical teams of the institute. The founding of IPHC also saw the creation of a team that inherited much of this expertise: the Accelerators' Instrumentation team. For many years, the team has devoted itself to design and development of beamlines, spectrometers and separators, diagnostics, the physics of charged particle beams, developments in electronics, computing and automation for large IN2P3 and CNRS projects.

## 2. Innovative accelerator projects

Within the first decade of 2000, three main projects arose in Europe with the ambition of building new facilities to further explore the frontiers of nuclear physics.

### 2.1. SPIRAL2

The SPIRAL2 project (GANIL, France) aims at producing Radioactive Ion Beam (RIB.) by ISOL as well as low-energy inflight methods. The SPIRAL2 facility [2], whose commissioning started in 2019, is composed of an injector including two ECR sources, a RFQ. operating at 88.05 MHz, a LINAC based on 88.05 MHz superconducting independently phase quarter wave superconducting cavities able to accelerate high intensity (5 mA) deuteron beam up to 20 MeV/u and light heavy ion (Q/A= 1/3) beam (1 mA) up to 14.5 MeV. The HEBT lines will distribute those beams to a beam dump or to the experimental stable ion beam experimental areas or transported to the 200 kW target ion source system. The expected rate of fission is $10^{14}$/s.

### 2.2. MYRRHA

The Multi-purpose hYbrid Research Reactor for High-tech Applications (MYRRHA) project is led by the SCK•CEN (Mol, Belgium). It aims at demonstrating the feasibility and operability of an efficient transmutation of nuclear waste products. MYRRHA will use external neutron produced by protons coming from Accelerator Driven Systems (ADS). The protons will have energy of 600 MeV and an intensity varying from 1 to 4 mA. The beam will be delivered to the reactor vertically from above through an achromatic beam line and window [3]. Currently under design, MYRRHA is foreseen to release first beams at 100 MeV in 2026.

### 2.3. FAIR

The Facility for Antiproton and Ion Research (FAIR), based at GSI (Darmstadt, Germany), has the aim of producing ions of all the natural elements in the periodic table as well as antiprotons [4]. FAIR will be composed of an accelerating ring of 1.1 km circumference (SIS100) connected to storage rings and experimental stations (including several km of beam lines). The overall facility should be completed in couple of years.

These projects require of being capable of mastering the quality of the beams with unprecedented efficiency. To manage this task, the use of reliable beam diagnostics is mandatory. It is in this context that IPHC proposed the development of a new device able to measure one of the parameters the most important when operating such beamlines.

## 3. Beam emittance measurement

The details of the transverse phase-space distribution is an important parameter when conceiving and operating accelerator beamlines. The so-called emittances give information on the size and the divergence of a beam. An emittance is the six-dimensional distribution of all position coordinates along the three configuration-space directions and the associated velocity coordinates. Typically, it is projected into the two-dimensional subsets { $x, x'$ }, { $y, y'$ } for the transverse emittances and { $z, z'$ } for the longitudinal emittance.

### 3.1. 2D emittances

The emittance of the beam along a transverse axis is called two-dimensional emittance as it provides information on the position and the angle along a single axis. Besides the Allison scanner that will be discussed in the following section, few 2D emittance measurement systems exist such as:

- The double-slit system. This emittance-meter consists of a uniaxial profile scanner of beamlets. It is made of two independent slits. The full beam impinges upon the first slit and is cut into small beamlets. After traversing a space of a certain length from the first slit, the profile of the small beamlet is measured by the second slit. For every position of the first slit there is a corresponding profile scan of second slit. This emittance-meter can be accurate, but its speed depends on the speed of the relative movement of the two slits.
- The slit-grid system. This principle of beamlet selection is the same as for the double-slit system [5, 6]. The beamlet propagates within a space without any field. The angular distribution is measured by a set of parallel wires. The angle of the beamlet is related to the position of the wire that collects it. The grid measures the current on each wire due to the emission of secondary electrons generated by the interaction of the beam with them. This system does not need any field to operate however its resolution is totally dependent on the center distance of the wires and the length of the space between the slit and the grid. Although very fine scanning can be achieved with this device, the mechanical development is complex and such device is rather expensive due to electronics that are needed and the positioning precision that is required.
- The quadrupole scan. In this method, the emittance value is obtained from the correlation between the beam size and the quadrupole magnetic field in the transformation matrix model [6, 7]. It involves minimal hardware to implement. A detector measuring the profile is placed at a certain distance from the quadrupole. This diagnostic principle is simple but based on several approximations including the absence of space charge because it assumes a linear perspective. Moreover, it is insensitive to geometric aberrations of the beam transport

### 3.2. 4D emittances

The possibilities of measuring 4D emittances are more limited than for 2D emittances. In this category, pepperpot is the main principle used [8, 9]. This interceptive device is based on the natural propagation of a beamlet within a distance. The selection of the beamlet is made by holes instead of slits. Indeed, to obtain the beam distribution in the 4D phase space and therefore the corresponding emittances, the measurements must be carried out on beamlets strategically distributed on the pepperpot plane, implying to use several selective holes [9]. The pepperpot allows measuring horizontal and vertical planes in a single-shot which makes this system faster that the methods described previously. While all the double slit methods just give the 2D projections x-x' and y-y' separately, Pepper-pot method gives a measure of the x-x'/y-y' correlations. However its accuracy is greatly limited by sampling. This is directly linked to the number of beamlets that are selected and therefore to the number of holes available to the pepper plane. In fact, this method does not describe the entire beam unlike 2D systems which scan the entire beam through successive selection. In the case of a pepperpot, only the parts of the beam selected by the sampler are analyzed.

### 4. Development of the Allison emittance-meter

In 1983, P.W. Allison proposed a scanner capable of measuring the trajectory angle distribution with an electric sweep while a mechanical scan probes the particle position distribution [10]. This makes the scan time shorter when compared to some of the principles discussed previously giving 2D emittances. This type of scanner allows obtaining the transverse emittances under the shape of ellipses even if the beam distribution presents distortion due to aberrations from the transport process. The presence of a rigid box with fixed slit positions minimizes the uncertainty from having two slits being misaligned. Moreover, as the Allison scanner does not use any wire, it cannot be impacted by any cross-talk generation. Therefore the increased signal-to-noise ratio allows for more effective halo analysis. The accuracy of the representation of the true two-dimensional particle distributions makes this type of scanner widely used [11-14]. This diagnostics has also the advantage of being robust to strong currents. Such device can offer total scanning times of 90 s for each plane if considering a beam of 30 mm diameter and measuring steps of 0.2 mm.

### 4.1. Operating principle

The system is a combination of an electric trajectory sweep technique together with a mechanical position beam sweep. Thus a simultaneous measurement of the position ($x$) and angle ($x'$) is performed for each position enabling the reconstruction of the full phase space display and calculation. When an ion with a charge q, a mass m and with an axial velocity, $v_z$, travels in an electrical field E its transverse position, $x$, changes (Fig. 1):

$$x = \int v_x \, dt = v_{x0}.t + \frac{q.E.t^2}{2.m} \qquad (1)$$

As

$$t = \frac{z}{v_z} = \frac{z}{\sqrt{\frac{2.q.U}{m}}}, \qquad (2)$$

This gives:

$$x = x'_0.z - \frac{\Delta V.z^2}{4.g.U} \qquad (3)$$

Where U is the ion potential.

For ions passing through the rear slit, when $x = 0$ and $z = L_{eff}$, within a deflection voltage, $\Delta V$, applied across a gap g, the angle of displacement, $x'_0$, is written:

$$x'_0 = \frac{\Delta V}{U} \frac{L_{eff}}{4.g} \qquad (4)$$

Also

$$x'_0 = \frac{\Delta V}{U} \frac{L_{eff}}{4.g} \cdot \frac{(L_{eff}+2\delta_2)}{(L_{eff}+\delta_1+\delta_2)} \qquad (5)$$

in the case of $\delta_1 \neq \delta_2$.

The maximum analysable angle, $x'_{max}$, is limited by the ions striking the deflecting angle plates when $x_{max} < \frac{g}{2}$, then:

$$x_{max} = x'_{max} \cdot \frac{L_{eff}}{2} - \frac{V.L_{eff}^2}{8.g.U} = \frac{L}{4} \cdot x'_{max} < \frac{g}{2} \qquad (6)$$

$$x'_{max} < \frac{2g}{L_{eff}} \qquad (7)$$

This corresponds to the maximum voltage, $\Delta V_{max}$:

$$\Delta V_{max} = \frac{8 \cdot g^2 \cdot U}{L_{eff}^2} \quad (8)$$

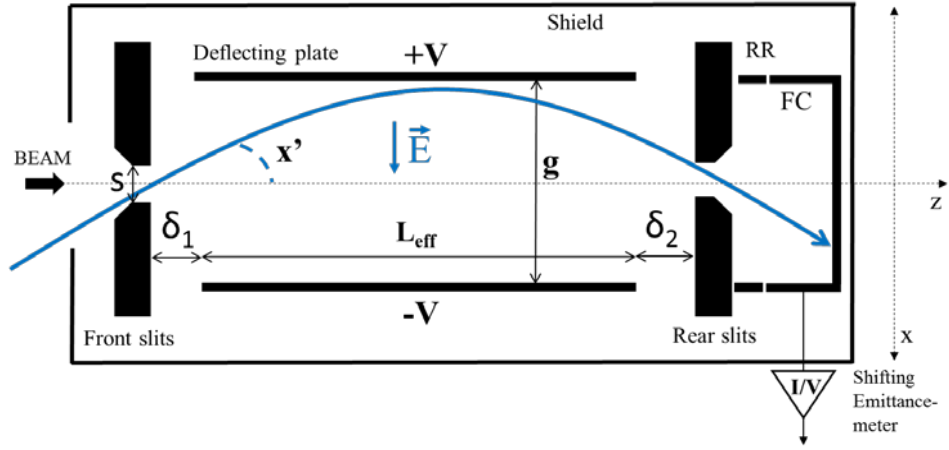

**Figure 1.** Schematic of the Allison principle. +V and –V are the potential of the plates. S is the aperture of the front slit. $L_{eff}$ is the effective length of the plates. $\delta_1$ and $\delta_2$ are the distances between the plates and the front and rear slits respectively. FC means Faraday Cup. g is the gap between the plates. E is the electrical vector induced by the voltage between the plates. RR is the electron repelling ring.

Figure 1 shows the schematic of the Allison principle. The scanning of the beam is done by moving up and down the device. The 2D scan process for each plan requires the scan of the plate voltage at each x-position. The stepping of the x-position aims at getting the profile of the divergence distribution.

From the distribution of particles, N, at coordinates x and x', the beam parameters can be calculated as following:

The average of the beam position, $\bar{x}$, and beam divergence, $\bar{x}'$, are defined by:

$$\bar{x} = \frac{1}{N} \sum_{i=1}^{n} x_i \text{ and } \bar{x}' = \frac{1}{N} \sum_{i=1}^{n} x'_i \quad (9)$$

Therefore the rms size, $\sigma_x$, and rms divergence, $\sigma_{x'}$, are written:

$$\sigma_x = \sqrt{\frac{1}{N} \sum_{i=1}^{n}(x_i - \bar{x})^2} \text{ and } \sigma_{x'} = \sqrt{\frac{1}{N} \sum_{i=1}^{n}(x'_i - \bar{x}')^2} \quad (10)$$

From there, the rms correlation xx' of the beam is determined by:

$$\sigma_{xx'} = \sqrt{\frac{1}{N} \sum_{i=1}^{n}((x_i - \bar{x}) \times (x'_i - \bar{x}'))} \quad (11)$$

The non-normalized emittance is:

$$\varepsilon_{xx'} = \sqrt{\sigma_x^2 \sigma_{x'}^2 - \sigma_{xx'}^4} \quad (12)$$

and the normalized emittance is calculated using the relativistic parameters of the beam, $\beta$ and $\gamma$:

$$\varepsilon_{xx'}^N = \beta \gamma \varepsilon_{xx'} \quad (13)$$

The Twiss parameters, $\alpha_{xx'}$, $\beta_{xx'}$ and $\gamma_{xx'}$ are defined as following:

$$\alpha_{xx'} = -\frac{\sigma_{xx'}^2}{\varepsilon_{xx'}}, \beta_{xx'} = -\frac{\sigma_x^2}{\varepsilon_{xx'}}, \gamma_{xx'} = -\frac{\sigma_{x'}^2}{\varepsilon_{xx'}} \tag{14}$$

From these, the equation of the rms ellipse of the beam is determined:

$$\gamma_{xx'}(x - \bar{x})^2 + 2\alpha_{xx'}(x - \bar{x})(x' - \overline{x'}) + \beta_{xx'}(x' - \overline{x'})^2 \tag{15}$$

Using the same method as for the xx' plane, the equation of the rms ellipse of the beam in the yy' plane can be written:

$$\gamma_{yy'}(y - \bar{y})^2 + 2\alpha_{yy'}(y - \bar{y})(y' - \overline{y'}) + \beta_{yy'}(y' - \overline{y'})^2 \tag{16}$$

### 4.2. Design of the emittance-meter

The Allison emittance-meter developed by IPHC is made of:
- Stainless steel deviation plates,
- Entrance and exit Tungsten slits,
- Electron suppressor made of bias-able ring capable of working up to 1 kV,
- Faraday Cup with a thermal screen made of copper,
- High voltage amplifier,
- High accuracy positioning system,
- I/V digitizer to read the beam current in the FC,
- Thermocouples.

The devices were designed to be able to measure normalized emittance ranges from 0.01 to 1 π.mm.mrad (or from 1.5 to 150 π.mm.mrad when non-normalized). The maximum angle that can be measured is 140 mrad (for a 30 keV beam) with a divergence resolution of 20 µrad. The normalized emittances of the beam in the LEBT and MEBT of SPIRAL2, MYRRHA and FAIR are found in the range from 0.2 to 0.3 π.mm.mrad.

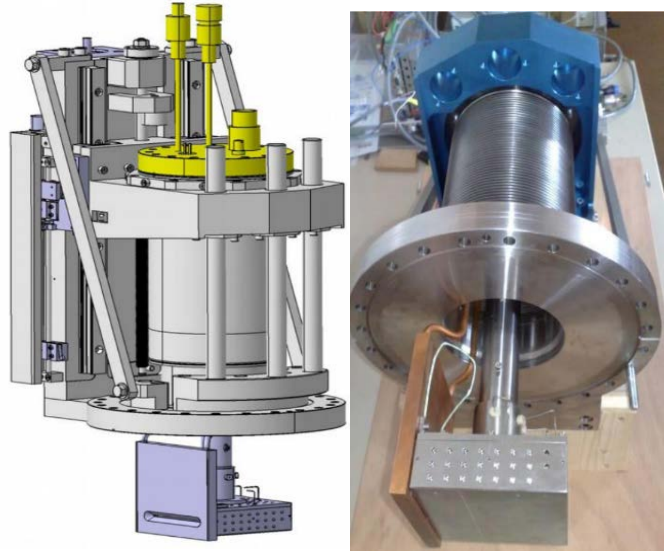

**Figure 2.** Technical drawing and picture of the IPHC emittance-meter.

The beam power met in the SPIRAL2 and FAIR projects are 400 W and 700 W respectively. For instance, in SPIRAL2 LEBT, diagnostics should support a power density of 20 W/mm$^2$ [15]. The scanner was designed to handle such power densities with a maximum of 32 to 40 W/mm$^2$.

The scanner itself composing the emittance-meter has typical dimensions up to 620x270x400 mm and can weigh up to 47 kg. Figure 2 shows the 3D drawing and the picture of the full device.

A total of two pairs of H/V Allison emittance-meters are used in SPIRAL2: two pairs in the LEBT and one in the MEBT. A third one was part of an Intermediate Test Bench (BTI) for the commissioning of the RFQ and the MEBT [16]. The BTI, presented in Fig.3, was composed of twelve beam diagnostics: BPMs, TOFs, profilers, CFR, BEM, emittance-meters, CF, slits and a Beam Dump.

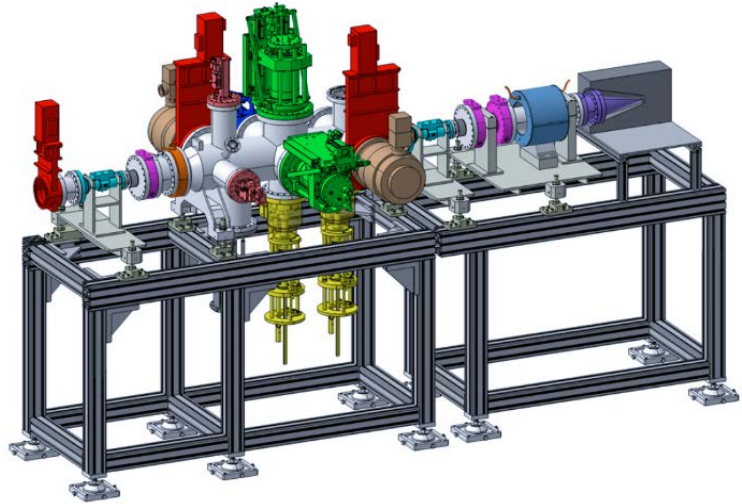

**Figure 3.** 3D view of the BTI.

Figure 4 shows the location of the two pairs in the LEBT of SPIRAL2 and one in the MEBT at the exit of the RFQ. A single device, located right after the RFQ, is used in MYRRHA [17] and a single one in the LEBT in the FAIR project.

Table 1 presents the different design parameters used for developing the Allison emittance-meters according to the projects. For SPIRAL2, the beam current impinges on the narrow slit of the scanner pod. The emerging beamlet passes between the pair of electric deflection plates driven by a linear ramp generator: 2.8 kV max (LBET) and 8 kV max (MEBT). The maximum measurable angles are ±100 mrad (LEBT) and ±30 mrad (MEBT).

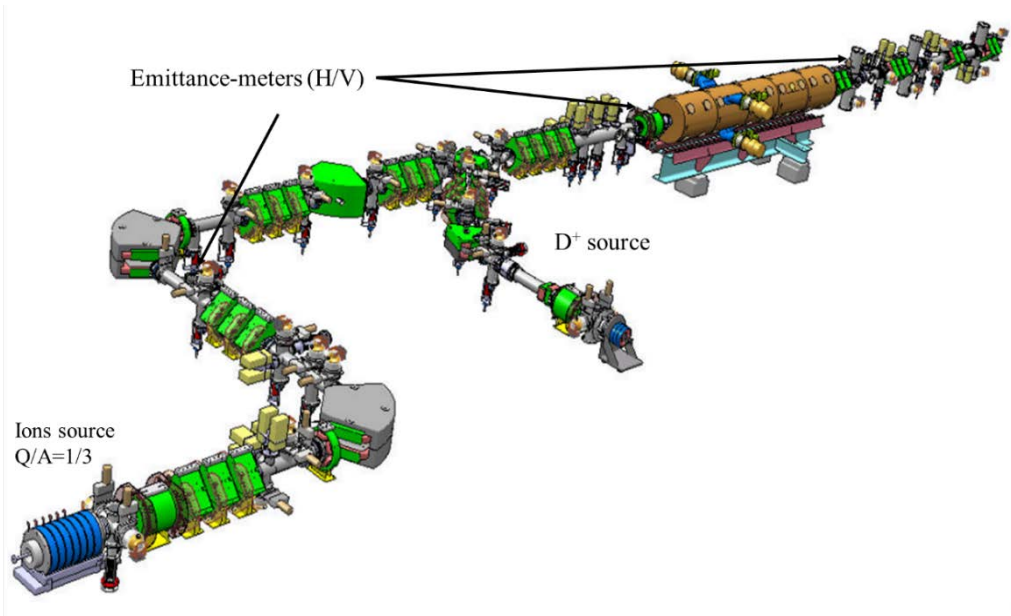

**Figure 4**. Location of the three pairs of IPHC emittance-meters on the LEBT and MEBT sections of SPIRAL2.

The scanner pod is stepped across the beam to complete the measurement as a function of $x$ and $x'$ (full stroke: 120 mm in the LEBT). The intensity of the beamlet is measured at each step of the pod. With x-slit (S in Table 1) and divergence slit widths of 0.12 mm, beamlets current is assumed to reach 10's µA for high intensity beam (few mA). In the frame of SPIRAL2, the divergence resolution, on the LEBT and MEBT is better than 30 µrad.

Table 1: Main design parameters versus projects

|  | SPIRAL2 (LEBT - MEBT) | MYRRHA (LEBT) | FAIR (LEBT) |
|---|---|---|---|
| Gap, G (mm) | 5 - 2 | 5 | 7 |
| Stroke (mm) | 120 - 120 | 120 | 180 |
| ΔV (kV) | 2.8 - 8.0 | 1.4 | 8.0 |
| S (mm) | 0.20 - 0.12 | 0.12 | 0.12 |
| $x'_{max}$ (mrad) | 100 - 30 | 30 | 30 |

To preserve the emittance-meter from any damage due to high beam intensity, the data collection requires operation at less than the full intensity of the injectors within the three facilities: ≃5.0 mA (SPIRAL2), ≃ 8.5 mA (MYRRHA) and ≃ 7.0 mA (FAIR).

Manipulating a beam of low energy may generate space charge effects leading to nonlinearity in the measurements. Several studies on the topic [18, 19, 20] shows that these effects can induce error up to tens percents in transverse emittance measurements of low energy beam by Allison Scanner (such as for H-, [21, 22]). Also, when doing the commissioning of the MYRRHA low energy beam transport, an increase of the gas pressure in the line could lower the emittances and the beam divergence resulting in a minimization of the nonlinear effects in the beam distribution [23]. Nevertheless, further investigations will be performed on the scanner developed at IPHC concerning these effects.

### 4.3. Command-Control System

The setup component for emittance measurement is distributed over three entities: a VME set under VxWorks, an API and a host station under Linux. The adopted software environment is EPICS (Experimental Physics and Industrial Control System) : the physical quantities to control, pilot or monitor are accessible via the internal Ethernet network by the EPICS Channel Access protocol. These instrumental quantities are represented by "records" in a "Database" hosted by the EPICS server commonly called IOC (Input Output Controller).

The VME equipment is composed of several cards: a processor card, an ADC (Analog to Digital Converter) card for reading temperatures and currents, a DIO (DIgital Inputs and Outputs) card used for hardware selections and configurations, a DAC (Digital to Analog Converter) card which generates the voltage ramps to be amplified, a high voltage generator card and a motor control card for the movements of the actuator with the measuring head. When the emittance measurement is requested, the program gets the user parameters and waits for validation of the PLC process. Then, the program applies the maximum voltages requested to the plates of the measurement head and comes in a loop made up of three sequences: positioning, voltage ramp setting and current reading.

The Graphical User Interface (GUI) dedicated to emittance-meters is developed in CSS-BOY (Control System Studio Best OPI Yet). In view of the complexity of the diagnosis, three levels of control and command are to be distinguished: data acquisition, configuration and debugging.
In the first case, only the maxima, minima, steps and gains are necessary for an emittance measurement. For an advance setting of the operation (such as timing, test currents, maximum temperature threshold, maximum time of exposure to the beam, etc.). As for the latter case, several panels are accessible to diagnose possible malfunctions of the various components of the emittance-meter (motors, VME cards, PLC, etc.). Figure 5, 6 and 7 show examples of view of the Graphical User Interface during an emittance measurement and an analysis.

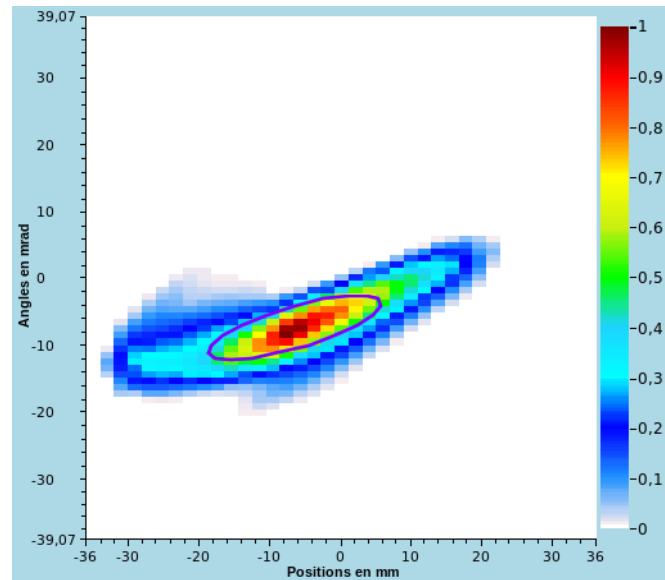

**Figure 5**. Example of emittance-meter measurement in x of a beam of $O^{6+}$ at the SPIRAL2 LBE beamline. The non-normalized emittance measured is 39.15 π.mm.mrad (60 keV beam kinetic energy).

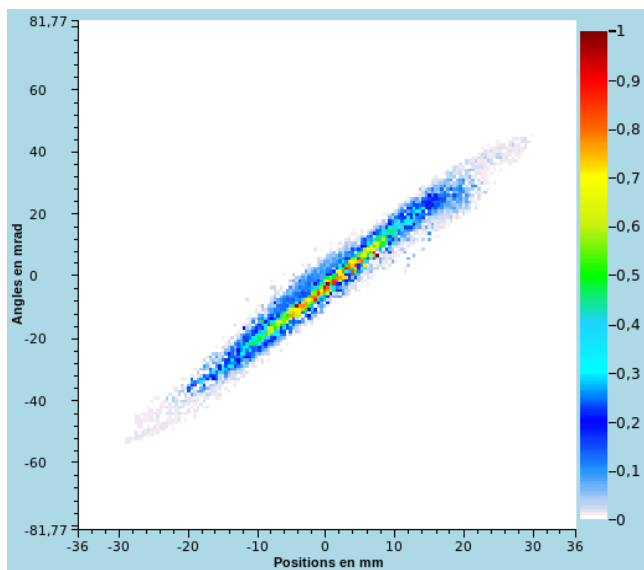

**Figure 6**. Example of emittance-meter measurement in y of a beam of H$^+$ at the SPIRAL2 LBE beamline. The non-normalized emittance measured is 27.77 π.mm.mrad (20 keV beam kinetic energy).

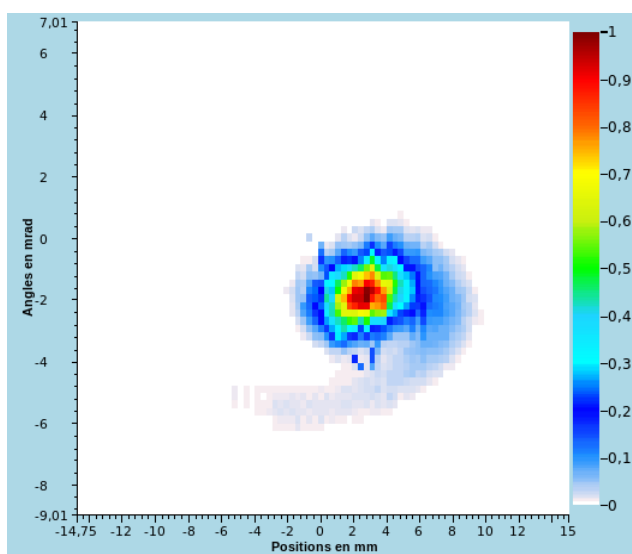

**Figure 7**. Example of emittance-meter measurement in y of a beam of H$^+$ at the SPIRAL2 LME beamline. The non-normalized emittance measured is 1.92 π.mm.mrad (730 keV beam kinetic energy).

Very good agreements were made between the values of emittances measured and the beam dynamics simulations as for example in SPIRAL2 [24].

## 5. Outcome

The success of the Allison emittance-meters developed at IPHC for SPIRAL2, MYRRHA and FAIR allowed agreements concerning a know-how license with the industry [25]. Further studies and improvements of the scanner are foreseen such as the real impact of the space charge effects in the measurements. Also, the development of new beam diagnostics are currently under investigations. The team will contribute to the development of diagnostics composing the future beamline of the new GANIL Injector [26].